\documentclass[5p,twocolumn]{elsarticle}
\usepackage{graphicx,latexsym}
\usepackage{dcolumn}
\usepackage{amssymb,amsmath,bm}
\usepackage{subfigure}
\usepackage{braket}
\usepackage{multicol}
\usepackage{siunitx}
\usepackage{array}
\usepackage{float}

\usepackage[nopar]{lipsum}
\usepackage{caption}
\usepackage[super]{nth}

\newcommand{\angstrom}{\text{\normalfont\AA}}
\usepackage{hyperref}
\hypersetup{
    pdfnewwindow=true,       
    colorlinks=true,         
    linkcolor=blue,          
    citecolor=blue,          
    filecolor=magenta,       
    urlcolor=black           
}

\usepackage[normalem]{ulem}

\def\sec#1{Sec.\ \ref{#1}}
\def\eq#1{Eq.\ (\ref{#1})}
\def\fig#1{Fig.\ \ref{#1}}
\def\tab#1{Tab.\ \ref{#1}}

\journal{}

\begin{document}

\begin{frontmatter}


\title{Properties of BSi$_6$N monolayers derived by first-principle computation}

\author[a1,a2]{Nzar Rauf Abdullah}
\ead{nzar.r.abdullah@gmail.com}
\address[a1]{Division of Computational Nanoscience, Physics Department, College of Science, 
             University of Sulaimani, Sulaimani 46001, Kurdistan Region, Iraq}
\address[a2]{Computer Engineering Department, College of Engineering, Komar University of Science and Technology, Sulaimani 46001, Kurdistan Region, Iraq}

\author[a1]{Hunar Omar Rashid} 

\author[a4]{Chi-Shung Tang}
\address[a4]{Department of Mechanical Engineering,
National United University, 1, Lienda, Miaoli 36003, Taiwan}

\author[a5]{Andrei Manolescu}
\address[a5]{Reykjavik University, School of Science and Engineering,
        	Menntavegur 1, IS-101 Reykjavik, Iceland}     

\author[a6]{Vidar Gudmundsson}   
\address[a6]{Science Institute, University of Iceland,
	Dunhaga 3, IS-107 Reykjavik, Iceland}

%

\begin{abstract}

The buckling effects due to BN-bonds in BN-codoped silicene, BSi$_6$N, on structural stability, electronic band structure, and mechanical, thermal and optical properties are studied systematically by first-principle calculations within density functional theory. In the presence of BN-bonds, a high warping in BSi$_6$N indicating a high buckling effect is found due to the presence of a repulsive interaction between B and N atoms. It thus breaks the sublattice symmetry of silicene and opens up a bandgap. The high buckling of BSi$_6$N leads to a decrease in its stiffness and thus induces fractures at small values of applied strain. The finite bandgap caused by the BN-bonds leads to enhancement of the Seebeck coefficient and the figure of merit, and induces a redshift of a peak in the dielectric response. 
By increasing the distance between the B and N atoms i.e. for the BSi$_6$N without BN-bonds, 
a flatter BSi$_6$N is found compared to pristine silicene. The stiffness of the structure and the ultimate strain are increased. The breaking of the sublattice symmetry is very weak and a very small bandgap is revealed. As a result, the Seebeck coefficient and the figure of merit stay very small. A reduction in the optical response is seen due to an indirect bandgap. 
\end{abstract}


\end{frontmatter}

\section{Introduction}

Two-dimensional (2D) materials are at the heart of nanotechnology because of their outstanding physical and chemical properties, and their potential applications in nanoelectronics \cite{Tao2015, LeLay2015}. 
Silicon structures indicating graphene-like silicene, a 2D material system exhibiting semi-metal properties caused by a zero bandgap, were experimentally investigated by Aufrey et al.\ \cite{doi:10.1063/1.3419932}. 
The first theoretical study of silicene was in 1994 using first-principles total-energy calculations \cite{PhysRevB.50.14916}. Afterwards, the electrical, thermal and optical properties of silicene 
have been extensivly investigated for understanding the structure. 
In fact, the zero bandgap limits the application of silicene in nanoelectronics despite of its high carrier mobility \cite{Pan2015}. Therefore, several techniques have been used to tune the bandgap to enhance the functionality of silicene in technologies \cite{ZHAO201624}, such as using an external electric field oriented perpendicular to the monolayer of Si atoms \cite{PhysRevB.85.075423}, bond symmetry breaking with uniaxial strain \cite{Jia2016}, surface adsorption \cite{Quhe2012}, alkali metal intercalation \cite{Liu_2014}, and substitutional doping of Boron- and Nitrogen-codoped silicene \cite{YIN201839}.

It has been found that silicene is relatively weaker than other 2D materials such as graphene in terms of stiffness and strength. In addition, it is more rigid when subject to bending, due to its slightly buckled molecular geometry \cite{ROMAN201450}. Silicene exhibits a strain-induced self-doping phenomenon, which is also closely related to its buckled structure, that cannot arise in graphene \cite{WANG20136}. Therefore, its mechanical properties are different compared to other 2D planar materials. It has been shown that a silicene lattice is stable up to $17\%$ under biaxial tensile strain \cite{doi:10.1063/1.4794812}.   
Qin et al.\ proposed that the in-plane stiffness of silicene is much weaker than that of graphene. So, the obtained strain is close to $20\%$ under uniform
expansion in ideal conditions \cite{doi:10.1063/1.4732134}. 
The nonlinear elasticity of silicene has been investigated and it has been demonstrated that the deformation and failure behavior, and the ultimate strength are all anisotropic \cite{C3RA41347K}.

Thermoelectric energy conversion in 2D materials such as silicene \cite{PhysRevB.88.115404} with impurities \cite{PhysRevB.89.165419,ABDULLAH20181432} has attracted tremendous interest due to significant potential for industrial applications. Theoretical studies of the electron and the phonon thermal conductivity have been reported \cite{PhysRevB.87.195417,Abdullah_2018}. Using equilibrium molecular dynamic simulations and the Kubo-Green method together, the thermal conductivity of silicene has been predicted to be $20$~W m$^{-1}$K$^{-1}$.
In non-equilibrium molecular dynamics simulations, the thermal conductivity of a monolayer silicene under uniaxial stretching is found to be around $40$~Wm$^{-1}$K$^{-1}$~\cite{PhysRevB.89.054310}. The Boltzmann Transport Equation has also been utilized to investigate the thermal properties of silicene 
and it indicated a conductivity about $9.4$~Wm$^{-1}$K$^{-1}$ at $300$~K. In all cases the thermal conductivity is much lower than for bulk silicon \cite{doi:10.1063/1.4870586}. 
Furthermore, the Seebeck coefficient and the figure of merit have been reported using first-principle density functional techniques and linear response in which
the higher figure of merit for distorted silicene is demonstrated \cite{PhysRevB.89.125403},
and for silicene nanoribbons it can be up to $160$ \cite{Sadeghi2015}. 
Recently, Boltzmann theory for electrons
under the relaxation time approximation has been employed to obtain the Seebeck coefficient 
showing that hydrogenation can greatly improve the electronic figure of merit, $ZT_e$, of multilayer silicene \cite{PhysRevB.99.235428}.

Another interesting aspect of silicene is its role in optical applications such as in the 
optoelectronic industry, in photo-detectors, sensors, low power lasers and not least in fiber optics communication \cite{doi:10.1021/acs.jpcc.6b08973,C8CS00338F}.
Optical response characteristics indicate that they strongly depend on the direction of polarization of the light \cite{gudmundsson2019coexisting}. The dielectric functions, and optical absorption are hence different for light polarized parallel, $E_{\parallel}$, and perpendicular, $E_{\perp}$, to the plane of silicence \cite{chinnathambi2012optical}. 
The optical response may also be changed by tuning the bandgap via a geometrical modification of a silicene nanosheet such as its width. As a result, a broad frequency photoresponse ranging from far infrared to ultraviolet is found \cite{C4RA03942D}. At low-frequency regime under an in-plane polarized driving field, the dynamical polarization, the dielectric function, and an absorption of a radiation in the infrared region have been obtained \cite{WU2018665}.

In this paper, we study the electronic, mechanical, optical, and thermal properties of silicene chemically modified by various Boron (B), and Nitrogen (N) atoms configurations doped in the silicene nanosheet structure, BSi$_6$N. In addition, the stability and formation energy calculated via density functional theory (DFT) are demonstrated. We show how different BN isomers that form BN-bonds influence the shape of BSi$_6$N via a buckling degree. Tuning the buckling degree caused by the BN-bonds  influences the entire physical properties of the system as will be discussed in the Results section.

In \sec{Sec:Model} the silicene structure is briefly over viewed. In \sec{Sec:Results} the main achieved results are analyzed. In \sec{Sec:Conclusion} the conclusion of the results is presented.

\section{Computational Tools}\label{Sec:Model}

Crystalline and molecular structure visualization programs (XCrySDen) and VESTA
are employed to visualize all the structures presented in this work \cite{KOKALJ1999176, momma2011vesta}.
After modeling the pristine silicene and BSi$_6$N structures, 
we use density functional theory employing the generalized gradient approximation (GGA) with the Perdew-Burke-Ernzerhof (PBE) functionals \cite{PhysRevLett.77.3865}. The
DFT calculations based on the standard Kohn–Sham (KS) equations are performed using the Quantum Espresso (QE) package \cite{Giannozzi_2009, giannozzi2017advanced}.
The Brillouin-Zone (BZ) sampling integrations have been achieved using sets of points
corresponding to a $14\times14\times1$ Monkhorst-Pack mesh \cite{PhysRevB.13.5188}.
The convergence of the mesh cut-off is checked and we find that at the value of $1088.45$ eV 
the GGA the model is converged with respect to our results. 
A geometry optimization is obtained with all residual forces 
being less than $10^{-6}$ eV/$\angstrom$. The distance between periodically
repeated images of the different systems along the $z$-direction is set to at least $20$~$\angstrom$.
In the density of state (DOS) calculations, a $77\times77\times1$ grid points mesh is used.

Bader charge analysis has been done using the Henkelman code for analysing the transfer 
rate of charge between atoms~\cite{Tang_2009}.
In addition, a Boltzmann transport theory (BoltzTraP) is used to study thermal properties of 
the systems \cite{madsen2006boltztrap-2}. The BoltzTraP code employs a mesh of band energies and 
is interfaced to the QE package \cite{ABDULLAH2020126578}. 
The optical properties of the systems are evaluated by the QE code.
An optical broadening of $0.1$~eV is assumed for the calculation of the dielectric properties.

\section{Results}\label{Sec:Results}

\begin{figure*}[htb]
 	\centering
 	\includegraphics[width=1.0\textwidth]{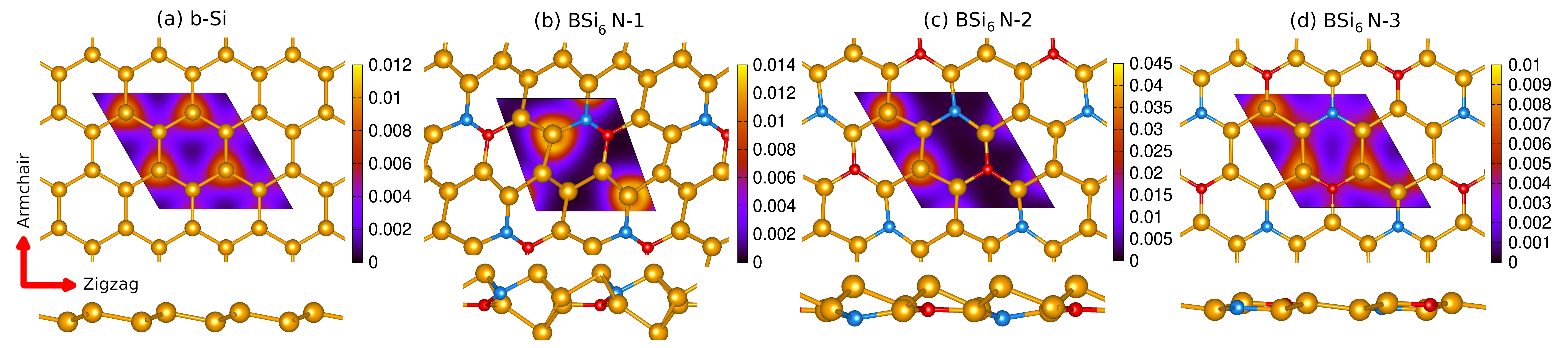}
 	\caption{Pristine buckled silicene (b-Si) (a), BN-codoped silicene identified as BSi$_6$N-1 (b), BSi$_6$N-2 (c), and BSi$_6$N-3 (d). The Si, B and N atoms are golden, blue and red colored, and the bottom panel is the side view of the corresponding structures. The contour plot is the electron charge distribution in a $2\times 2$ supercell.}
 	\label{fig01}
 \end{figure*}

In this section, we present our model, the calculation of the formation energy, the electronic properties including the band structure, the mechanical, the thermal, and the optical properties 
of pristine silicene and the BSi$_6$N structure.
We assume a $2\times2$ supercell of buckled silicene structure presented in \fig{fig01}(a).
The buckled nature of silicene is caused by the positions of the Si atoms in the A and B sites that do not lie in the same plane leading to a staggered sublattice potential and a layer separation between 
the two sublattices of A and B sites.
In addition to pure buckled silicene (b-Si), three BN-codoped silicene structures (three isomers) are considered where the B atom (blue) is fixed at a para position at the top of the hexagon \cite{RASHID2019102625}.
In isomer b, the B and N (red) are at the ortho positions lead to BN-bonds, in isomer c same type of dopants (B or N) are placed at the adjacent positions, in isomer d, B and N are at the para positions \cite{ABDULLAH2020103282}. 
The isomer b, c, and d are identified as BSi$_6$N-1, BSi$_6$N-2, BSi$_6$N-3, respectively. 
The numbers of 1, 2, and 3 are linked to BSi$_6$N for distinguishing these three structures. 
After fully relaxation of b-Si, the hexagonal lattice constant, a,  the buckling length, $\delta$, and the Si-Si bond length of pristine b-Si are $3.86$~$\angstrom$, $0.446$~$\angstrom$, $2.26$~$\angstrom$, respectively, (see \tab{table_one}).
\begin{table}[h]
	\centering
	\begin{center}
		\caption{\label{table_one} Lattice constant, a, buckling length, $\delta$, Si-Si, B-N, Si-B, and Si-N bonds for all b-Si and BN-codoped structures. The unit of all parameters is $\angstrom$.}
		\begin{tabular}{|l|l|l|l|l|l|l|}\hline
			Structure	  & a     & $\delta$& Si-Si   & B-N     & Si-B    &   Si-N   \\ \hline
			b-Si	      & 3.86  &  0.446  & 2.26    & -       & -       &  -       \\
			BSi$_6$N-1	  & 3.41  &  1.301  & 2.38    & 1.377   & 2.064   &  1.779   \\
			BSi$_6$N-2	  & 3.37  &  0.682  & 2.22    & -       & 1.92    &  1.831    \\
			BSi$_6$N-3	  & 3.42  &  0.281  & 2.25    & -       & 1.87    &  1.81    \\   \hline
		\end{tabular}
	\end{center}
\end{table}
These values are in a good agreement with previous studies of slightly buckled silicenes \cite{PhysRevLett.102.236804}. The lattice constants, the buckling lengths, and the bond lengths 
of BSi$_6$N structures are also listed in~\tab{table_one}.

 \lipsum[0]
We can see from \tab{table_one} that the B and N atoms configurations influence the lattice constant, and the buckling lengths. The bond lengths in turn affect the sublattice symmetry of the system and the bandgap. In presence of BN-bonds, in BSi$_6$N-1, the buckling degree is strongly affected and the structure records the highest buckling length, $d = 1.301$~$\angstrom$. As a result, the structure is strongly warped as is shown in the side view (bottom panel) of \fig{fig01}(b). 

If the B and N atoms are doped in silicene in such a way that one Si atom is located between these two atoms (see \fig{fig01}(c)), the buckling length of BSi$_6$N-2 gets smaller, $d = 0.682$~ $\angstrom$, or the distortion becomes smaller comparing to BSi$_6$N-1. 
Increasing further the distance between the B and N atoms, see \fig{fig01}(d), the structure is close to the pristine b-Si structure or even flatter as $d = 0.281$~$\angstrom$.
The warping and flattening of these structures are attributed to the interaction between the B and N atoms. This issue will be discussed later. 

The contour plots in \fig{fig01} represent the electron charge distributions in which the charge density of those atoms near to the surface is high.  The charge distribution of BSi$_6$N-1 indicates that the electron densities are not delocalized along the B-N bond. This suggests a repulsive interaction between B and N atoms \cite{aliofkhazraei2016graphene}. On the other hand, a
Bader charge analysis demonstrates that an N atoms in BSi$_6$N-3 gain more charge from the surrounding
Si atoms as the repulsive interaction between the B and N atoms does not play an important role there. Therefore, a charge accumulation around the N atoms is observed.


\subsection{Structural stability} 

The isomers in which the BN-bonds are present (ortho positions), 
have considerably larger cohesive energy and are hence more stable than their
counterparts (para positions) in which the B and the N atoms are apart.
This is due to the more stability of the B–N bond than the B–Si or the N–Si bonds.
On the other hand, the formation energy can be used to find the most stable structure.
The formation energy, $E_f$, of a doped system is defined in terms of the structural energy of a cell containing N$_i$ atoms of species $i$, 
\begin{equation}\label{eq01}
E_f = E_{\rm T} - N_{\rm Si} \, \mu_{\rm Si} - N_{\rm B} \, \mu_{\rm B} -  N_{\rm N} \, \mu_{\rm N}   
\end{equation}
where E$_{\rm T}$ is the total energy of the BN-codoped silicene structure, N$_{\rm Si}$,  N$_{\rm B}$, and N$_{\rm N}$ are the number of Si, B and N atoms, respectively, and $\mu_{\rm Si}$,  $\mu_{\rm B}$, and $\mu_{\rm N}$, are the chemical potential of the Si, B and N atoms, respectively. 
The chemical potential is determined by the energy per atom. 
Using \eq{eq01} the formation energy of BSi$_6$N-1, BSi$_6$N-2, BSi$_6$N-3 are $-122.16$, $-120.59$ and $-121.05$~eV, respectively, indicating that the BSi$_6$N-1 is energetically the most stable structure among these three structures.

\subsection{Interaction energy}

The interaction energy between the B and N atoms doped in the silicene structure can be calculated 
from the total energy. In BN-codoped graphene, the interaction between the B and the N atoms  indicates that B-B and N-N interactions in the system are repulsive whereas the N–B interaction is attractive \cite{aliofkhazraei2016graphene}. In our BN-codoped silicene, the Bader charge analysis indicates the repulsive interaction between B and N atoms as both doped atoms are positively charged.
In addition, the interaction energy in BN-codoped graphene is inversely varied with the distance between B and N atoms, and the interaction strength is almost zero when the separation distance is greater than or equal to $4.0$~$\angstrom$ \cite{doi:10.1063/1.4742063}. So, we expect that the interaction between B and N atoms in silicene is also inversely proportional to the distance between these two atoms.

\lipsum[0]
\begin{figure*}[htb]
	\centering
	\includegraphics[width=0.65\textwidth]{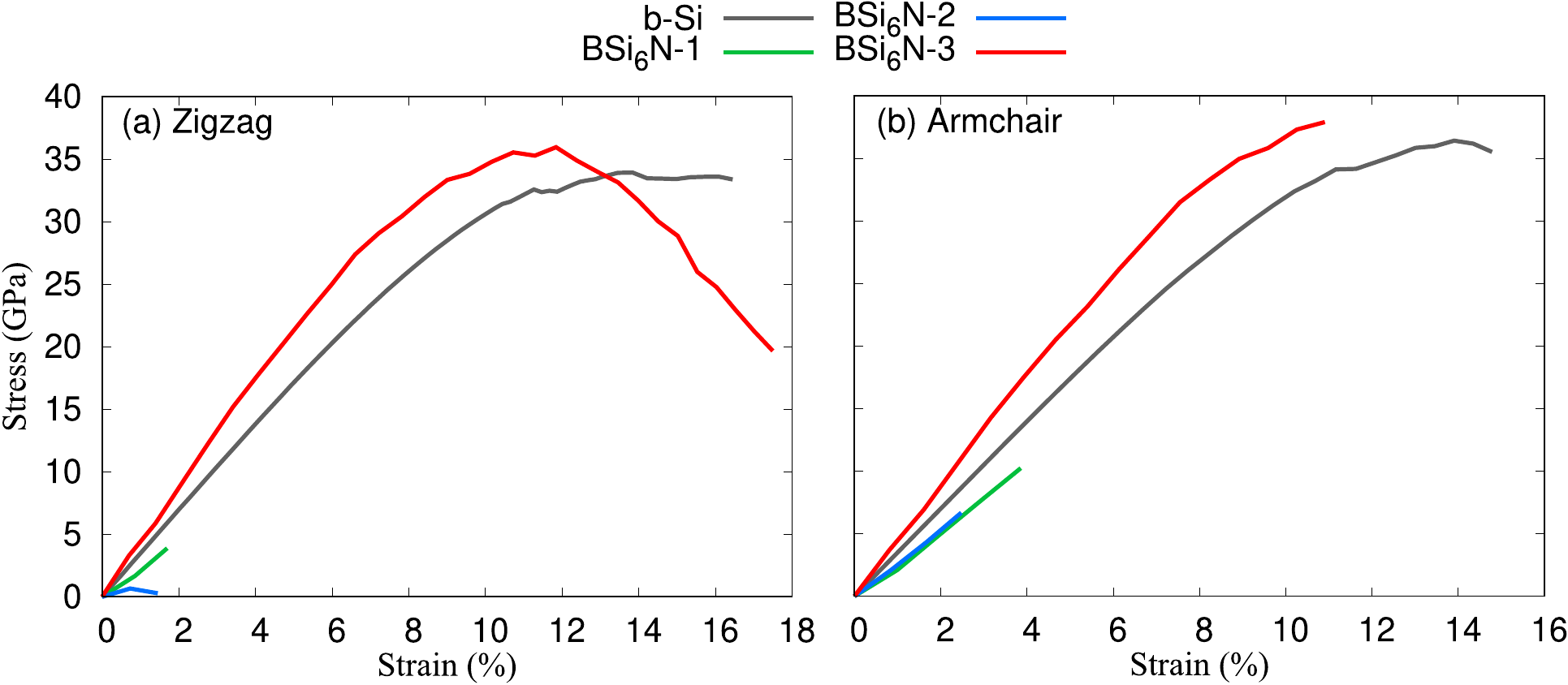}
	\caption{Stress-strain curves of pure b-Si (gray), BSi$_6$N-1 (green), BSi$_6$N-2 (blue), and BSi$_6$N-3 (red) for zigzag (a) and armchair (b) directions. The systems are subjected to uniaxial strain.}
	\label{fig02}
\end{figure*}

After fully relaxing our systems, we found that the distance between the B and N atoms in BSi$_6$N-1, 
BSi$_6$N-2, and BSi$_6$N-3 is $1.377$, $3.428$, and $4.085$~$\angstrom$, respectively. Furthermore, the total energy obtained from the SCF calculations indicates that the interaction between B and N atoms in BSi$_6$N-1 is maximum while in BSi$_6$N-3 is almost zero \cite{ABDULLAH2020126350}. 
This is the reason that the BSi$_6$N-1 and BSi$_6$N-3 are the most warped and flattened structures, respectively.

\subsection{Mechanical properties}

The DFT calculations can be utilized to study the mechanical properties of single-layer and freestanding silicene. 
Uniaxial tensile simulations are carried out to probe the stress-strain properties. 
This can be obtained by gradually applying load in the zigzag or armchair directions of a structure \cite{MORTAZAVI2017228}.
During uniaxial tensile loading, the periodic dimension
along the loading direction is increased step-by-step with a fixed
strain of $0.02$.

The stress-strain curves for three
BSi$_6$N structures are plotted together with the one for the
b-Si in \fig{fig02} for the $x$-, the zigzag, (a) and the $y$-, the
armchair direction (b).
We first focus on the pure b-Si structure (gray). The first observation is that the buckling length of the sheets gradually decreases by increasing strain level. A linear relation between the stress and strain reflecting an increase in the in-plane stiffness and elastic property is seen.
The linear elastic regime for the pure b-Si ends at $\approx 10\%$ (gray) in both zigzag and armchair directions revealing isotropic elastic response.
Higher strain leads to a stretching of the Si-Si bonds and, therefore, a decrease in the 
in-plane stiffness.  
The ultimate strength of pure b-Si structure in the zigzag direction is $33.92$~GPa at strain $13.8\%$ while in the armchair direction it is $36.42$~GPa at strain $13.9\%$.
As we have mentioned before, silicene is relatively weak structure comparing to other 2D materials such as graphene due to its buckled geometry. We therefore expect that the ultimate strength of silicene is lower than that of graphene \cite{Nzar_arXive_2020, aliofkhazraei2016graphene, suzuki2010silicene}.

In the BSi$_6$N structures, the interaction between B and N atoms and BN-bonds play an essential role on stress-strain curve. Both BSi$_6$N-1 (green), and BSi$_6$N-2 (blue) induce fracture strain very early at strain less than $1.7\%$ in the zigzag direction and $4\%$ in the armchair direction.
The small BN-bond length and the larger Si-Si bond in BSi$_6$N-1 comparing to b-Si, and  
the relatively strong repulsive interaction between the B and N atoms in BSi$_6$N-1 induces a strong deformation and larger buckling length. A higher buckling length, and a less in-plane stiffness is obtained. Therefore, the fracture strain of both BSi$_6$N-1 and BSi$_6$N-2 is small.

The stress-strain curve of the last structure, BSi$_6$N-3 (red), is interesting and completely different from BSi$_6$N-1 and BSi$_6$N-2 because the repulsive interaction between B and N atoms is almost zero. The elastic region here is decreased to $\approx 6.5\%$  in the zigzag and 
$\approx 7.5\%$ in the armchair directions indicating anisotropic mechanical properties of the structure in the elastic region. 
The ultimate strength is $35.94$~GPa at strain $11.84\%$ and $37.9$~GPa at strain $10.9\%$, in the zigzag and armchair directions, respectively. The ultimate strength of BSi$_6$N-3 in both direction is higher than that of pure b-Si. So BSi$_6$N-3 is mechanically the stronger structure.

\subsection{Electronic properties}

\begin{figure*}[htb]
	\centering
	\includegraphics[width=0.9\textwidth]{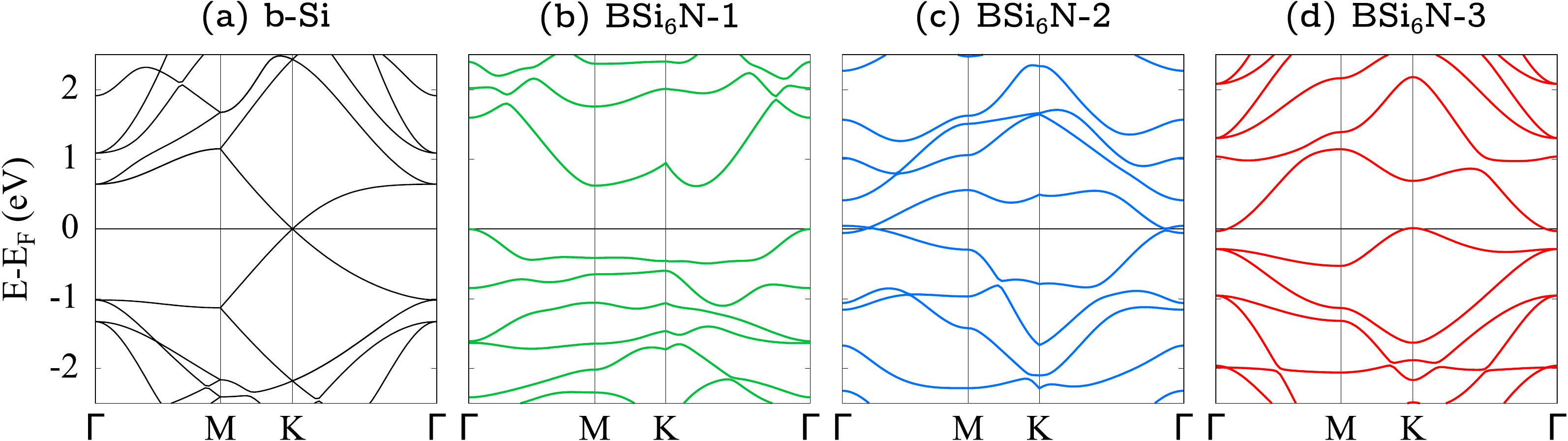}
	\caption{Band structure for optimized structures of b-Si (a), BSi$_6$N-1 (b), BSi$_6$N-2 (c), and BSi$_6$N-3 (d). The energies are with respect to the Fermi level, and the Fermi energy is set to zero.}
	\label{fig03}
\end{figure*}

In this section, we discuss the electronic band structure of pure silicene and BSi$_6$N using DFT. 
In \fig{fig03}, we plot the band structure of b-Si (a), BSi$_6$N-1 (b), BSi$_6$N-2 (c), and BSi$_6$N-3 (d) along the high symmetry points in the Brillouin zone. 
The linear dispersion of b-Si band structure around the Fermi level indicates the semimetallic nature where valence band maxima ($\pi$ band) and the conduction band minima ($\pi^*$ band) touch each other only at the high symmetry K point. The linear dispersion in pure b-Si shows that the charge carriers around the Fermi level will act like massless Dirac-Fermions. 
The Hamiltonian that introduces the electronic structure of b-Si around the Dirac point can be written as \cite{aliofkhazraei2016graphene}
\begin{equation}
 \hat{H} = 
\begin{pmatrix}
	\Delta               & \hbar v_F(k_x-i k_y) \\
	\hbar v_F(k_x+i k_y) & \Delta
\end{pmatrix}
\end{equation}

\lipsum[0]

Herein, $\Delta$, $v_F$, and $k$ refer to onsite energy difference between the Si atoms located at A and B sites, Fermi velocity, and  momentum  of charge carriers. 
The linear relation refers to the zero value of the onsite energy difference, $\Delta$, in b-Si arising from the presence of inversion symmetry in b-Si.

In BSi$_6$N structures, the band structure is importantly modified as a bandgap is
opened and tuned depending on the B and N atoms configurations as is shown in \fig{fig03}(b-d).
In general, a band gap in BSi$_6$N is opened near the K point due to breaking of
the inversion symmetry by the distortion generated by the B and N atoms configuration. 
The reason for the broken symmetry is that the potential seen by the atoms at sites A and B is now different, leading to a finite value of onsite energy $\Delta \neq 0$, where  $\Delta = \alpha (V_{\rm A} - V_{\rm B})$ with $\alpha$ being a constant value and V$_{\rm A}$(V$_{\rm B}$) is the potential seen by an atom at the site A(B). On the other hand, the potential difference is directly proportional to the buckling parameter, $V_{\rm A} - V_{\rm B} \sim \delta$, where $\delta$ is the buckling parameter  \cite{Rahman_2014}. 

In BSi$_6$N-1 (b), the buckling parameter, $\delta$, has the highest value among three BN-codoped structures leading to an opening of a bandgap and the linear dispersion at K point is not seen anymore. 
The BSi$_6$N-2 (c) has an intermediate buckling length and the opening bandgap at the K point is still observed but the Fermi energy slightly crosses the valence band maxima near the $\Gamma$ point leading to degenerate semiconductor behavior of BSi$_6$N-2. 
This is very much consistent with study of N-doped silicene reported in \cite{PhysRevB.87.085444}. 
In the last structure, BSi$_6$N-3 (d), the buckling parameter is low but the potential difference between A and B sites leads to an indirect bandgap along the K and the $\Gamma$ points.

\subsection{Thermal properties}

The study of thermal properties of pure silicene and BSi$_6$N sheets is important for thermoelectric energy conversion which is the ability of a device to convert temperature gradient into an electrical current. The figure of merit, $ZT$, reflecting the efficiency of a device and the Seebeck coefficient, $S$, are calculated here. The figure of merit is defined as $ZT = \sigma S^2 T/\kappa$, where $\sigma$ is the electric conductance, $T$ is the absolute temperature, and $\kappa = \kappa_e + \kappa_p$ is the total thermal conductance that consists of the electron, $\kappa_e$ and the phonon, $\kappa_p$ contributions. 
Silicene is thermally stable up to $1500$~K which is attributed to a low phonon conductance. This is contrary to other 2D materials such as graphene, which exhibits a high thermal conductivity.

\begin{figure}[htb]
	\centering
	\includegraphics[width=0.35\textwidth]{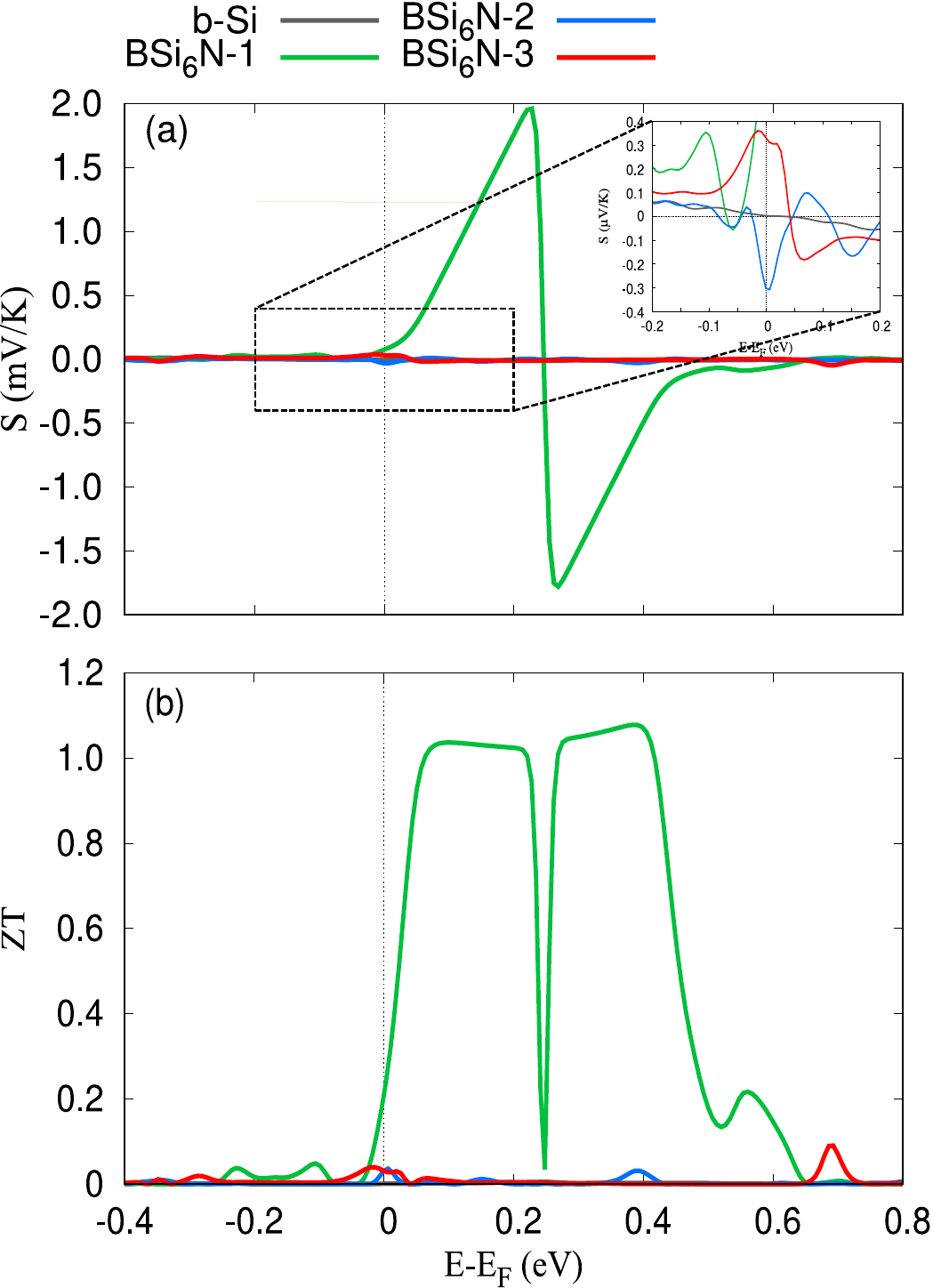}
	\caption{Seebeck coefficient (a), and the figure of merit (b) as a function of energy are plotted at $T = 100$~K for pristine b-Si (gray), BSi$_6$N-1 (green),  BSi$_6$N-2 (blue), and BSi$_6$N-3 (red).}
	\label{fig04}
\end{figure}

We show the thermal properties in the temperature range of $T = 20 \text{-} 160$~K where the
electron and the lattice temperatures are decoupled and the energy exchange between the charge carriers and the acoustic phonons is very weak \cite{Gabor648, PhysRevB.87.035415, ABDULLAH2018199,abdullah2019thermoelectric}. As a result, only the electronic part of the thermal properties is dominant in this temperature range. To calculate the electronic thermal properties of a material, the BoltzTraP software is used.

\lipsum[0]
\begin{figure*}[htb]
	\centering
	\includegraphics[width=0.8\textwidth]{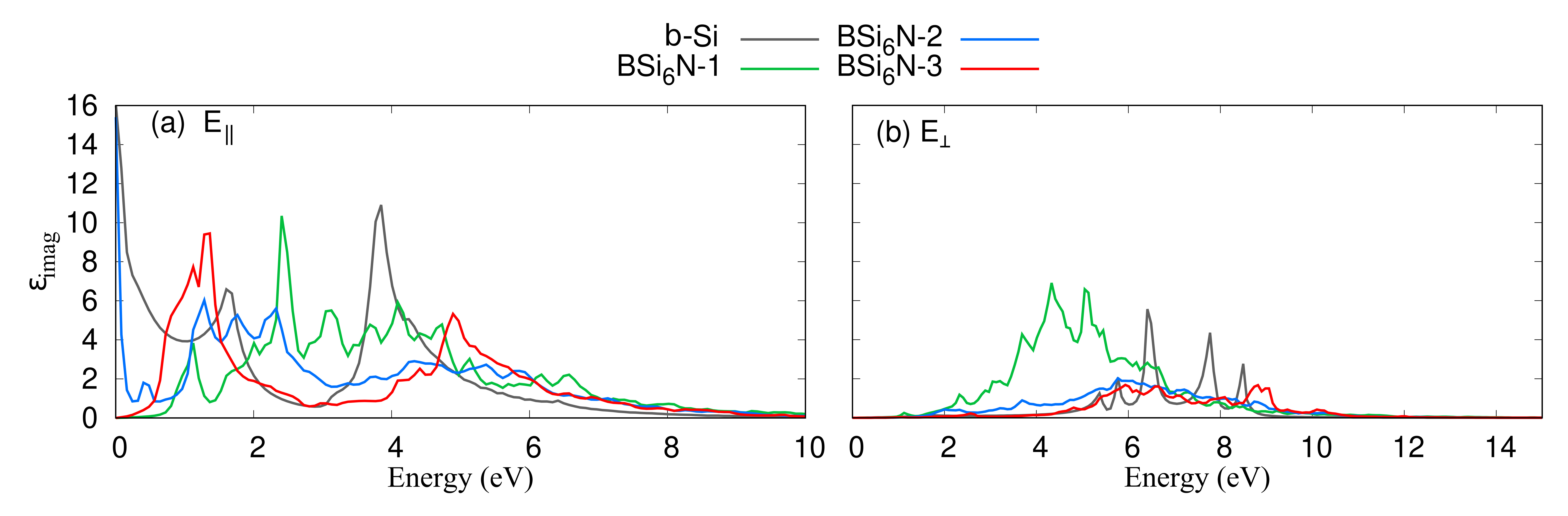}
	\caption{Imaginary part of dielectric function of pristine b-Si (gray), BSi$_6$N-1 (green),  BSi$_6$N-2 (blue), and BSi$_6$N-3 (red) for parallel, $E_{\parallel}$, (a) and perpendicular, $E_{\perp}$, (b) polarized electric field.}
	\label{fig05}
\end{figure*}

The Seebeck coefficient (a) and the figure of merit (b) versus energy at $T = 100$~K are shown in \fig{fig04} for pristine b-Si (gray), BSi$_6$N-1 (green),  BSi$_6$N-2 (blue), and BSi$_6$N-3 (red). 
It can clearly be seen that both $S$ and and $ZT$ show large variations for BSi$_6$N-1. 
This is attributed to the presence of a larger bandgap in BSi$_6$N-1. The strong dip in ZT of BSi$_6$N-1 is expected as the S is zero at the same energy, $0.25$~eV.
The inset in \fig{fig04}a indicates $S$ in the energy range of -2.0 to 2.0 eV. It is clear that the $S$ for b-Si is smaller than BN-codoped silicene structures. 

This is expected as the thermoelectric power in semiconductors of n and p types is approximately three orders of magnitude higher than that of metals. 
The change of sigh of the Seebeck coefficient indicates the transition between electron and hole dominated transport. 
In the n-type semiconductor, BSi$_6$N-1, there will be high thermoelectric figure of merit if there is a large energy difference between the conduction band minima and the Fermi energy, $E_{\rm c}\text{-}E_{\rm F}$, while for the p-type semiconductor $E_{\rm F}\text{-}E_{\rm v}$ is important for similar behavior where $E_{\rm v}$ is the valence band maxima.

\subsection{Optical properties} 

The optical properties of a material show how it interacts with electromagnetic field of an incident light \cite{Abdullah_2019, Abdullah2019}. Optical characteristics are directly related to electronic band structure. So, tuning the bandgap of a material will influence its optical behavior.

To see the influences of B and N doping of silicene on its optical response, we present the imaginary part of the dielectric function, $\varepsilon_{\rm imag}$, in \fig{fig05} for light polarized parallel (a) and perpendicular (b) to the plane of the b-Si (gray), BSi$_6$N-1 (green),  BSi$_6$N-2 (blue), and BSi$_6$N-3 (red). The imaginary part is related to the absorption of energy within the material. In order to obtain a more accurate optical response, we consider a finer large mesh size of k-points, $110\times 110\times 1$. It can be seen that the imaginary part of dielectric function changes with the direction of light polarization as the parallel and perpendicular components of $\varepsilon_{\rm imag}$ are different. Another observation is that the $\varepsilon_{\rm imag}$ in parallel polarized light is dominant below the energy range of $5$~eV while in the case of perpendicular polarized of light $\varepsilon_{\rm imag}$ is observed to be active above $5$~eV. This anisotropy in $\varepsilon_{\rm imag}$ has been reported and it is a consequence of the 2D
nature of the buckled silicene sheet~\cite{Chowdhury_2016}. 
In pristine b-Si, two major peaks are found in energy range from $0$ to $5.5$~eV in the case of parallel polarized light. 
The first weaker peak is located at $1.68$~eV indicating transition from $\pi$ to $\pi^*$ states which are close to the Fermi energy. The second peak is found at $3.85$~eV corresponding to the $\sigma$ to $\sigma^*$ transitions. It should be mentioned that the shape of both peaks is almost symmetric which is related to the symmetry of $\pi$ to $\pi^*$ bands, and $\sigma$ to $\sigma^*$ bands. 
Similar to graphene, inter-band transitions for perpendicular polarized light, $E_{\perp}$, are observed for pristine b-Si except the transitions here occure below $10$~eV (see \fig{fig05}(b) gray color).

In the BSi$_6$N structures, a redshift in the first peak corresponding to $\pi \rightarrow \pi^*$ transitions in the case of $E_{\parallel}$ is seen. The energy value of redshifted first peak is found to be $1.12$, $1.28$, and $1.36$~eV for BSi$_6$N-1,  BSi$_6$N-2, and BSi$_6$N-3, respectivily. The redshift can be referred to the opening of a bandgap along the M-K paths. In addition, the reduction in second peak for BSi$_6$N-1, and BSi$_6$N-2 in the case of $E_{\parallel}$ refers to the deformation and a further separation of the $\sigma$ and $\sigma^*$ bands arising from the warped BSi$_6$N-1, and BSi$_6$N-2 structures. But a blueshift and reduction in the second peak of BSi$_6$N-3 is seen which stands for the larger energy spacing between the $\sigma$ and $\sigma^*$ states. This is attributed to the smaller buckling parameter in BSi$_6$N-3. It has been reported that at a higher buckling degree a smaller energy spacing between $\sigma$ and $\sigma^*$ is obtained \cite{yan_2014}. This is the reason why the energy spacing in band structure of planer silicene is higher than that of buckled silicene. The same senario can be applied to the main transition peaks for $E_{\perp}$.

\lipsum[0]
\begin{figure*}[htb]
	\centering
	\includegraphics[width=0.8\textwidth]{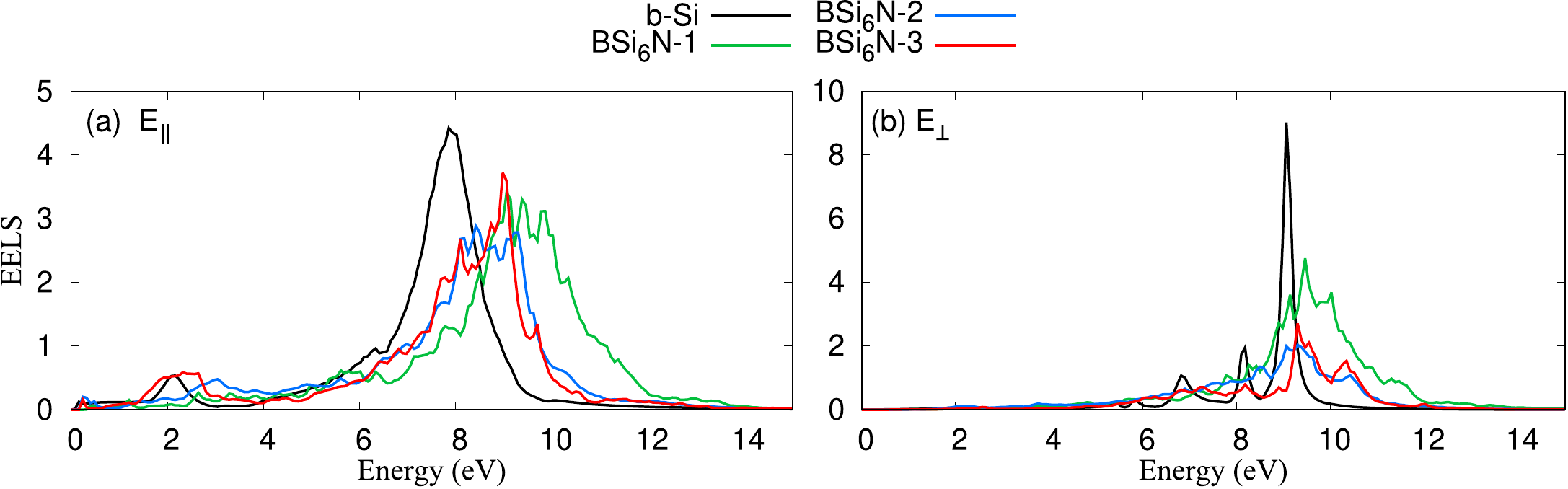}
	\caption{Electron energy loss spectroscopy (EELF) of pristine b-Si (gray), BSi$_6$N-1 (green),  BSi$_6$N-2 (blue), and BSi$_6$N-3 (red) for parallel (a) and perpendicular (b) polarized electric field.}
	\label{fig06}
\end{figure*}

The last study of our work is the electron energy loss function (EELF) which introduces the energy loss of a fast moving electron in a material. The EELF is shown in \fig{fig06} for 
b-Si (gray), BSi$_6$N-1 (green),  BSi$_6$N-2 (blue), and BSi$_6$N-3 (red) in the case of $E_{\parallel}$ (a) and $E_{\perp}$ (b). The peaks in ELLF indicate the point of transition from metallic to dielectric property of a material (the plasma frequency). In pristine b-Si, three peaks around $2.1$, $5.9$, and $7.95$~eV are observed for $E_{\parallel}$, and the same peaks with a shift towards a higher energy are seen at $6.82$, $8.17$, and $9.06$~eV for $E_{\perp}$. 
The first peak at $2.1$~eV reveals a $\pi$ plasmon, and the second peaks around $6$~eV stand for a $\pi+\sigma$ plasmon. The presence of these peaks is due to collective excitations at different light energy. The peak position of b-Si in both directions of light polarization are in a good agreement with previous studies of b-Si \cite{JOHN2017307}. 
The peaks in the ELLF corresponds to the dip in the complex dielectric function in both directions of light polarization. We therefore see shifting in the ELLF peaks of all three BSi$_6$N structures.

\section{Summary and Conclusions}\label{Sec:Conclusion}

Density functional theory was utilized for investigating the electronic characteristics of BSi$_6$N sheets, where the B and N atomic configuration and BN-bonds play a major role in opening bandgaps. It was shown that the repulsive interaction between the B and N atoms gives rise a high buckling degree in the system, which is reflected in a strongly broken sub-lattice symmetry of the system. The repulsive interaction between the B and N atoms is decreased by increasing distance between these two doped atoms. The high buckling degree in the presence of BN-bonds in BSi$_6$N reveals a weakness of the structure.
As a result, the resistance to deformation in response to an applied strain is decreased. 
In addition, the opening of a bandgap in the presence of the BN-bonds increases the Seebeck coefficient and the figure of merit. This may be interesting in high thermoelectric efficiency nanodevices.
A redshift towards low energy in the optical responses of BSi$_6$N has been found that may be  
interested in the regard of optoelectronic devices.  

\section{Acknowledgment}
This work was financially supported by the University of Sulaimani and 
the Research center of Komar University of Science and Technology. 
The computations were performed on resources provided by the Division of Computational 
Nanoscience at the University of Sulaimani. CST acknowledges support from Ministry of Science and Technology in Taiwan, under grant No. MOST 109-2112-M-239-003.
 


\end{document}